\documentclass[onecolumn,manuscript,showpacs,amsmath,apsrev4-1]{revtex4-1}

\usepackage{}
\usepackage{txfonts}
\usepackage{pifont}
\usepackage{amssymb}

\usepackage{dcolumn}
\usepackage{amsmath}
\usepackage[dvips]{epsfig}

\makeatletter
\def\btt#1{\texttt{\@backslashchar#1}}%
\DeclareRobustCommand\bblash{\btt{\@backslashchar}}%
\makeatother

\begin{document}
\title{Band-filling and correlation controlling electronic properties and magnetism
in K$_{x}$Fe$_{2-y}$Se$_{2}$: A slave boson study}
\author{Da-Yong Liu$^{1}$, Ya-Min Quan$^{1}$, Xiao-Jun Zheng$^{1}$, Xiang-Long Yu$^{1}$, Liang-Jian Zou$^{1,
        \footnote{Correspondence author, Electronic mail:
           zou@theory.issp.ac.cn}}$
}
\affiliation{ \it $^1$ Key Laboratory of Materials Physics,
Institute of Solid State Physics,
Chinese Academy of Sciences, P. O. Box 1129, Hefei 230031,
People's Republic of China
\\}
\date{\today}

\begin{abstract}
In this paper we investigate the electronic and magnetic properties of
K$_{x}$Fe$_{2-y}$Se$_{2}$ materials at different band fillings utilizing
the multi-orbital Kotliar-Ruckenstein's slave-boson mean field approach.
We find that at three-quarter filling, corresponding to KFe$_{2}$Se$_{2}$,
the ground state is a paramagnetic bad metal.
Through band renormalization analysis and comparison with the
angle-resolved photoemission spectra
data, we identify that KFe$_{2}$Se$_{2}$ is also an intermediate
correlated system, similar to iron-pnictide systems.
At two-third filling, corresponding to the Fe$^{2+}$-based systems,
the ground state is a striped antiferromagnetic (SAFM) metal with spin
density wave gap partially opened near the Fermi level.
In comparison, at half filling case, corresponding to the Fe$^{3+}$-based
compounds, besides SAFM, a $N\acute{e}el$ antiferromagnetic metallic ground
state without orbital ordering is observed in the intermediate correlation
range, and an orbital selective Mott phase (OSMP) accompanied with an
intermediate-spin to high-spin transition is also found.
These results demonstrate that the band filling and correlation control
the electronic state, Fermi surface topology and magnetism in
K$_{x}$Fe$_{2-y}$Se$_{2}$.

\end{abstract}

\pacs{74.70.Xa,71.27.+a,71.10.-w}
\maketitle

\section{INTRODUCTION}

Recently a new iron selenide superconductor K$_{x}$Fe$_{2-y}$Se$_{2}$
with $\it{T}_{\rm{c}}$ above 30 K \cite{PRB82-180520} has
attracted considerable attention for its unique
high $N\acute{e}el$ transition temperature and insulating properties,
as well as the presence of intrinsic Fe-vacancy ordering
\cite{EPL94-27009,PRB83-140505R,CPL28-086104,PRL107-137003},
quite different from the other iron-based superconducting materials.
These unusual properties arise an assumption that K$_{x}$Fe$_{2-y}$Se$_{2}$ may be a strongly correlated system.
%
KFe$_{2}$Se$_{2}$ is isostructural to the 122 system, {\it e.g.} BaFe$_{2}$As$_{2}$,
but chemically is close to FeSe.
On average there are 6.5 electrons with equal ratio of Fe$^{2+}$ and Fe$^{+}$
in KFe$_{2}$Se$_{2}$, rather than 6 ones in the iron-pnictide parent compounds
with solely Fe$^{2+}$.
Therefore it can be regarded as an electron overdoped 11 system,
in contrast to the underdoped KFe$_{2}$As$_{2}$ which has 5.5 electrons with equal
ratio of Fe$^{2+}$ and Fe$^{3+}$, and FeAs which has 5 ones with Fe$^{3+}$.
As a consequence, only the electron Fermi surface (FS) pockets exist
around the $M$ points in KFe$_{2}$Se$_{2}$, as observed in
recent angle-resolved photoemission spectroscopy (ARPES) experiments
\cite{nmat10-273,PRL106-187001} and in electronic
structure calculations \cite{PRB84-054502,CPL28-057402}.
Thus, the FS nesting between the hole pocket around the $\varGamma$ point and the
electron pocket around the $M$ point, which widely exists in FeAs-based materials,
is absent in KFe$_{2}$Se$_{2}$ compound.
%

Due to its unique electronic structure properties,
the magnetic properties of K$_{x}$Fe$_{2-y}$Se$_{2}$ are focused on.
Because of the difficulty of single crystal preparation for
pure AFe$_{2}$Se$_{2}$ (A=K, Tl, Rb, or Cs), its magnetic properties are
mainly studied theoretically, but still a debating issue, as addressed in what follows:
the local density approximation (LDA) calculations suggested that KFe$_{2}$Se$_{2}$
is a striped antiferromagnetic (SAFM) order,
same to the 1111 and 122 phases of the FeAs-based materials \cite{CPL28-057402};
while some others \cite{PRB84-054502} thought it to be bi-collinear AFM with
($\pi$/2, $\pi$/2) wave-vector, similar to FeTe \cite{PRL102-247001}.
However, the LDA calculations suggested that TlFe$_{2}$Se$_{2}$ is a checkerboard AFM
($\pi$, $\pi$) order \cite{PRB79-094528}, and the dynamical spin susceptibility
obtained within random phase approximation (RPA) also suggested the ($\pi$, $\pi$)
instability in KFe$_{2}$Se$_{2}$ \cite{physicab407-1139,PRB83-100515R}.
These discrepant results show that further investigations on the magnetism are deserved to
understand the unique properties in AFe$_{2}$Se$_{2}$ (A=K, Tl, Rb, or
Cs) compounds.
%

%
Many recent experiments  \cite{nphys8-126,SR2-212,PRX1-021020,PRB83-184521,
PRB85-094512,SR2-221,arXiv1203.1533} reported the wide existence of the
phase separation in K$_{x}$Fe$_{2-y}$Se$_{2}$ materials.
A scanning tunneling spectroscopy (STS) experiment demonstrated the
phase-separated component KFe$_{2}$Se$_{2}$ is the parent phase
contributing to the superconductivity, and K$_{0.8}$Fe$_{1.6}$Se$_{2}$
component is an Fe-vacancy order insulator \cite{nphys8-126},
implying that pure KFe$_{2}$Se$_{2}$ in the normal state
is more possibly a paramagnetic (PM) phase.
%
One may notice that different K contents in K$_{x}$Fe$_{2-y}$Se$_{2}$
lead to different band fillings of Fe 3$d$ orbitals,
hence to quite different electronic and magnetic properties.
Further Chen {\it et al.} reported that the electronic states
and magnetic phase diagrams of K$_{x}$Fe$_{2-y}$Se$_{2}$ system are closely
connected with the Fe valences \cite{SR2-212}.
We also notice that various theoretical magnetic configurations obtained within
the first-principles calculations for AFe$_{2}$Se$_{2}$ (A=K, Tl, Rb, or
Cs) do not include the Coulomb correlation correction
\cite{CPL28-057402,PRB84-054502,PRB79-094528}, which
implies a weak electronic correlation in AFe$_{2}$Se$_{2}$,
in contrast with the intermediate electronic correlation in the
FeAs-based compounds \cite{JPCM24-085603}.
Thus a few of questions are urgent to be answered: what is the realistic
electron filling ? Moreover how does the band filling affect
the electronic properties and magnetism in K$_{x}$Fe$_{2-y}$Se$_{2}$ ?
Considering that KFe$_{2}$Se$_{2}$ is a possible parent phase of superconducting
state, we will focus on K$_{x}$Fe$_{2-y}$Se$_{2}$ system in the absence
of ordered Fe vacancy at various electron fillings throughout this paper.

%
In this paper, to uncover the role of electronic correlation on the groundstate
properties of AFe$_{2}$Se$_{2}$, we adopt Kotliar-Ruckenstein's slave boson (KRSB)
mean field approach \cite{PRL57-1362,JPCM24-085603,EPJB85-55} to study the magnetic and
electronic properties at different band fillings in K-doped iron
selenides. Based on our previous LDA calculation results \cite{physicab407-1139},
we first present an effective three-orbital model for KFe$_{2}$Se$_{2}$,
and then determine the ground states of this model at different electron fillings.
We show that the ground state of K$_{x}$Fe$_{2-y}$Se$_{2}$
at fillings of 3/4, 2/3 and a half is a PM metallic phase, a SAFM with
orbital ordering, and a $N\acute{e}el$ AFM one without orbital ordering
in the intermediate and strong correlation regimes in addition to an orbital
selective Mott phase (OSMP) related with an intermediate-spin to high-spin transition,
respectively, showing that the band filling controls not only the
FS topology, but also the electronic structure and magnetic properties
of Fe-based superconducting materials.
The rest of this paper is organized as follows: a three-orbital tight-binding model
and the multi-orbital slave-boson mean-field approach are presented in {\it Sec. II};
the numerical results and discussions are
shown in {\it Sec. III}; the last section is devoted to the remarks
and summary.

\section{Three-orbital tight-binding model and slave boson approach}

Based on our previous electronic structure calculations \cite{physicab407-1139},
we find that FS is mainly
contributed by three $\it{t}_{\rm{2g}}$ orbitals,
thus the system can be described by a three-orbital model,
similar to iron pnictides \cite{PRB78-144517,PRB79-054504,
PRB79-064517,PRB81-014511}.
We extract an effective three-orbital tight-binding
model from our LDA band structures.
The tight-binding model Hamiltonian for the three-orbital model
in the momentum space is described as,
\begin{eqnarray}
   H_{0} &=&
   \sum_{\substack{k,\alpha,\beta,\sigma}}(\epsilon_{\alpha}\delta_{\alpha\beta}+
   T^{\alpha\beta}(\textbf{k})) C_{k\alpha\sigma}^{\dag}C_{k\beta\sigma}-
   \mu\sum_{k\alpha\sigma}n_{k\alpha\sigma},
\end{eqnarray}
where $T^{\alpha\beta}(\textbf{k})$ is the kinetic energy term,
$\epsilon_{\alpha}$ denotes the on-site energy of the $\alpha$
orbital, and $\mu$ is the chemical potential determined
by the electron filling.
The three-orbital tight-binding fitting of the
Fe-3$\it{d}$ bands is displayed in the solid lines, in comparison with the
original five bands \cite{physicab407-1139} in the dot lines, as shown in Fig. 1.
It is obviously found that the band structures
in KFe$_{2}$Se$_{2}$ are similar to that of LaFeAsO \cite{PRL105-096401},
only the position of Fermi energy $E_{F}$ is shifted.
Therefore this model can describe both the FeSe-based
and FeAs-based systems through changing the chemical potential.
\begin{figure}[htbp]
\hspace*{-9mm}
\begin{minipage}[t]{0.45\linewidth}
    \centering
    \includegraphics[width=3.3in]{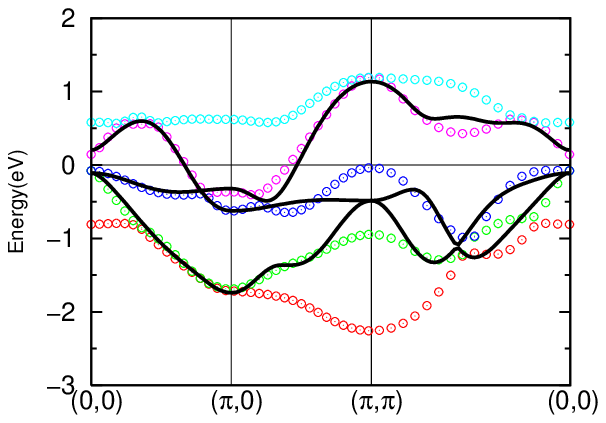}
    \label{Fig1-1}
\end{minipage}
\hspace{0.25ex}
\begin{minipage}[t]{0.45\linewidth}
    \centering
    \includegraphics[width=2.5in]{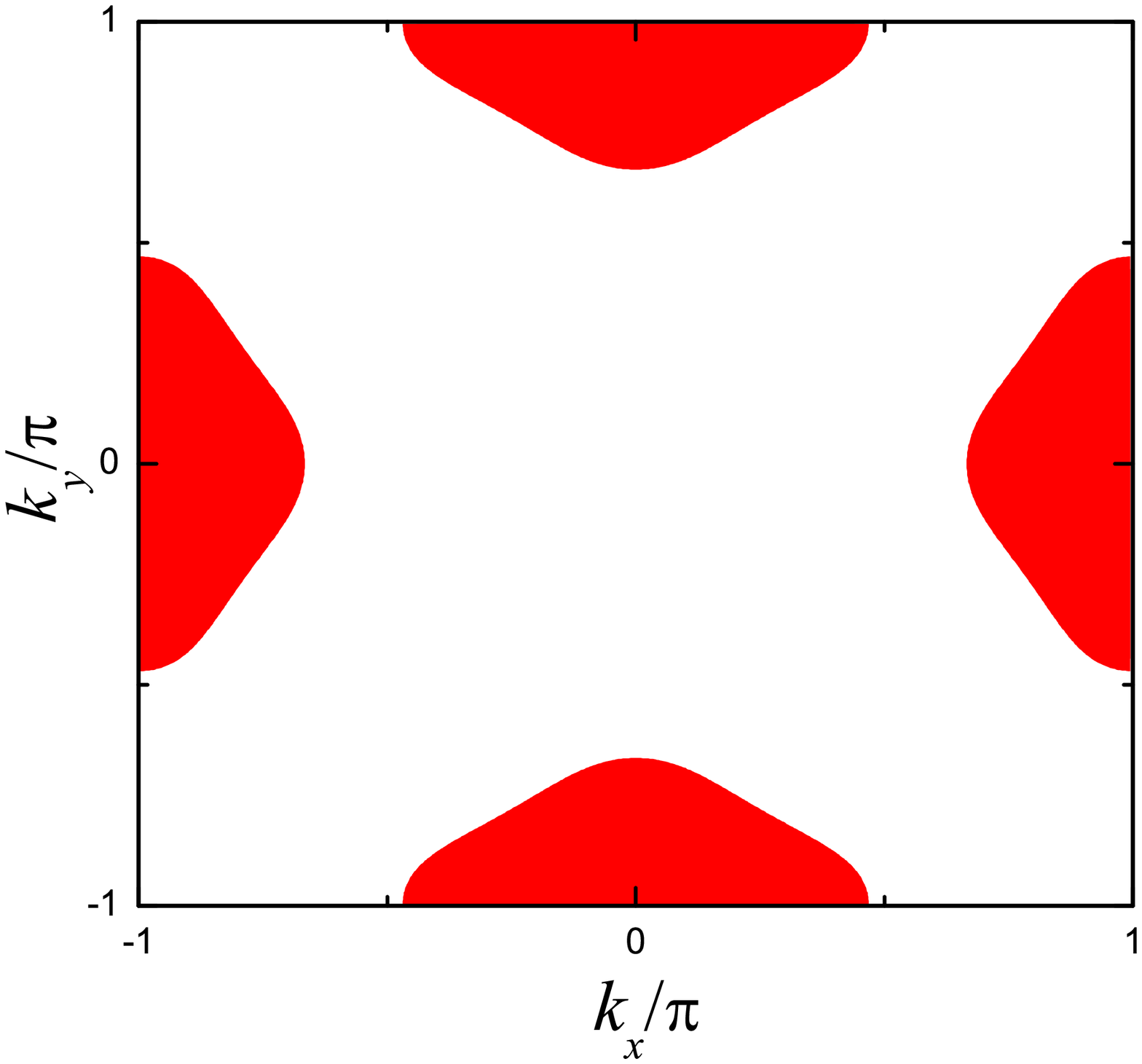}
    \label{Fig1-2}
\end{minipage}
\  \\
\caption{(Color online) Left panel: the band structures of
the Fe-3$\it{d}$ orbitals obtained by the LDA and its three-orbital tight-binding fitting.
Dot lines are the LDA five bands \cite{physicab407-1139}, and solid lines the fitted effective three bands.
Right panel: the corresponding Fermi surface for parent material
KFe$_{2}$Se$_{2}$ at the first Brillouin zone obtained by the three-orbital tight-binding model.}
\end{figure}

The tight-binding parameters of the three-orbital model for
KFe$_{2}$Se$_{2}$ are listed in the following.
The on-site energies measured from the Fermi energy for the
three orbitals are ($\epsilon_{1},\epsilon_{2},\epsilon_{3}$)=
($-$511.92, $-$511.92, $-$341.63), respectively, in units of meV.
Here orbital indices (1, 2, 3) indicate the $\it{d_{xz}}$,
$\it{d_{yz}}$, and $\it{d_{xy}}$ components, respectively.
Similar to BaFe$_{2}$As$_{2}$ \cite{PRB81-214503},
the orbital-dependent kinetic energy matrix elements are
expressed in terms of the inter-obrital and intra-orbital
hopping integrals as follows,
\begin{eqnarray}
  T^{11/22} &=& 2t_{x/y}^{11}cos k_{x}+2t_{y/x}^{11}cos k_{y}+4t_{xy}^{11}cos k_{x}cos k_{y} \nonumber \\
  &&\pm 2t_{xx}^{11}(cos 2k_{x}-cos 2k_{y})+4t_{xxy/xyy}^{11}cos 2k_{x}cos k_{y} \nonumber \\
  &&+4t_{xyy/xxy}^{11}cos 2k_{y}cos k_{x}+4t_{xxyy}^{11}cos 2k_{x}cos 2k_{y}, \nonumber
\end{eqnarray}
\begin{eqnarray}
  T^{33} &=& 2t_{x}^{33}(cos k_{x}+cos k_{y})+4t_{xy}^{33}cos k_{x}cos k_{y}+2t_{xx}^{33}(cos 2k_{x}+cos 2k_{y}) \nonumber \\
  &&+4t_{xxy}^{33}(cos 2k_{x}cos k_{y}+cos 2k_{y}cos k_{x})+4t_{xxyy}^{33}cos 2k_{x}cos 2k_{y}, \nonumber
\end{eqnarray}
\begin{eqnarray}
  T^{12}&=& 4t_{xy}^{12}sin k_{x}sin k_{y}+4t_{xxy}^{12}(sin 2k_{x} sin k_{y}+sin 2k_{y} sin k_{x}) \nonumber \\
  &&+4t_{xxyy}^{12}sin 2k_{x}sin 2k_{y}, \nonumber
\end{eqnarray}
\begin{eqnarray}
  T^{13/23}&=& \pm 2it_{x}^{13}sin k_{x/y}\pm 4it_{xy}^{13}cos k_{y/x} sin k_{x/y} \nonumber \\
  &&\pm 4it_{xxy}^{13}cos k_{y/x} sin 2k_{x/y}, \nonumber
\end{eqnarray}
The intra-orbital and inter-orbital hopping parameters
up to the fifth nearest-neighbor for the fitting of the three-band
structure in Fig. 1 are shown in Table {\it I}.
\begin{table}[htbp]
\begin{center}
\caption{The nonzero matrix elements of intra-orbital $t_{i}^{\alpha\alpha}$ and
inter-orbital $t_{i}^{\alpha\beta}$ hopping parameters up to fifth neighbors
of the three-orbital tight-binding model through fitting
the band structures. All the parameters are in units of meV.}
 \vspace{0.2cm}
\begin{tabular}{lcccccccccccccc}
\hline \hline
&$t_{i}^{\alpha\alpha}$ &$i = x$       &$i = y$     &$i = xy$   &$i = xx$     &$i = xxy$     &$i = xyy$   &$i = xxyy$  \\
\hline
&$\alpha$ = 1/2         &60.87     &77.65   &21.83  &54.35    &$-$34.35  &11.97   &31.67      \\
\hline
&$\alpha$ = 3             &$-$53.39  &        &301.2  &21.61    &$-$31.6   &        &$-$70.04        \\
\hline \hline
&$t_{i}^{\alpha\beta}$  &$i = x$     &$i = xy$      &$i = xxy$      &$i = xxyy$     \\
\hline
&$\alpha\beta$ = 12       &        &$-$119.57 &50.99       &$-$12.55        \\
\hline
&$\alpha\beta$ = 13/23     &302.39  &122.5     &$-$14.64    &            \\
\hline \hline
\end{tabular}
\end{center}
\end{table}
%
%

Considering the Coulomb interaction, in addition
to the kinetic term in Eq. (1), we describe the electronic
interaction part of the multi-orbital Hamiltonian as follows,
\begin{eqnarray}
  H_{I} &=& U\sum_{\substack{i, \alpha}}n_{i\alpha\uparrow}n_{i\alpha\downarrow}
  +U^{'}\sum_{\substack{i\\ \alpha\ne\beta}}n_{i\alpha\uparrow}n_{i\beta\downarrow}
  +(U^{'}-J_{H})\sum_{\substack{i,\sigma\\ \alpha<\beta}}n_{i\alpha\sigma}n_{i\beta\sigma}
\nonumber\\
  &&-J_{H}\sum_{\substack{i\\ \alpha\ne\beta}}
  C_{i\alpha\uparrow}^{\dag}C_{i\alpha\downarrow}C_{i\beta\downarrow}^{\dag}C_{i\beta\uparrow}
  +J_{H}\sum_{\substack{i\\ \alpha\ne\beta}}
  C_{i\alpha\uparrow}^{\dag}C_{i\alpha\downarrow}^{\dag}C_{i\beta\downarrow}C_{i\beta\uparrow}
\end{eqnarray}
where $\it{U}$($\it{U}^{'}$) denotes the intra-(inter-)orbital Coulomb
repulsion interaction and $\it{J}_{\rm{H}}$ the Hund's
rule coupling. Considering the rotation symmetry of the system,
we adopt $\it{U}^{'}$=$\it{U}$-2$\it{J}_{\rm{H}}$.

We notice that in FeAs-based materials, the electronic filling for the present three-orbital model
is only at two-third filling, {\it i.e.} $\it{n}$=4, the correct
FS can be reproduced \cite{PRB79-064517,PRB81-014511,PRL105-096401}.
In KFe$_{2}$Se$_{2}$ compound,
the Fe ions have two kinds of valence, Fe$^{2+}$ and Fe$^{+}$ with equal ratio,
and the average electron number is 6.5 per site,
different from Fe$^{2+}$ with 6 electrons in FeAs-based parent materials.
Consequently, pure KFe$_{2}$Se$_{2}$ should be at a filling of three quarters
({\it i.e.} $\it{n}$=4.5) at which
the FS can be reproduced correctly in the present three-orbital model.
The obtained FS of KFe$_{2}$Se$_{2}$ is plotted at $k_{z}$=0 in Fig. 1,
with four electron-like FS pockets.
The hole-like FS pockets at $\varGamma$
are absent. Such a FS topology is in agreement with the recent ARPES experiments \cite{nmat10-273,PRL106-187001} and the band structure
calculations \cite{PRB84-054502,CPL28-057402}.
To explore the band filling dependence of the electronic state, we also extend
the band filling to 2/3 and half, which corresponds to the Fe$^{2+}$ and
Fe$^{3+}$ systems, respectively.

KRSB mean field approach is known as
one of the effective methods to treat the wide electronic correlation
in the many-body systems.
Here, we extend the single-orbital KRSB mean-field method to the multi-orbital situation \cite{JPCM24-085603,EPJB85-55} and apply it on K$_{x}$Fe$_{2-y}$Se$_{2}$.
We introduce new fermion operators $\it{f_{i\alpha\sigma}}$ slaved by boson operators $\it{e_{i}}$, $\it{p_{i\alpha\sigma}}$, $\it{d_{i\alpha\sigma_{\beta}\sigma_{\gamma}}}$,
$\it{b_{i\alpha}}$, $\it{t_{i\alpha\beta\sigma}}$, $\it{r_{i\sigma_{\alpha}\sigma_{\beta}\sigma_{\gamma}}}$,
$\it{q_{i\alpha}}$, $\it{u_{i\alpha\sigma_{\beta}\sigma_{\gamma}}}$, $\it{v_{i\alpha\sigma}}$ and $\it{s}$,
which represent the empty,
single occupation with $\alpha$ orbital and $\sigma$ spin,
double occupation with $\beta$ orbital spin
$\sigma_{\beta}$ and $\gamma$ orbital spin $\sigma_{\gamma}$,
double occupation with two electrons in orbital $\alpha$,
triplicate occupation with
two electrons in orbital $\alpha$ and one electron in orbital $\beta$ with
spin $\sigma$,
triplicate occupation with each electron in orbital $\alpha$, $\beta$ and $\gamma$
with spin $\sigma_{\alpha}$, $\sigma_{\beta}$, and $\sigma_{\gamma}$,
quaternity occupation with two electrons in orbital
$\beta$ and $\gamma$,
quaternity occupation with two electrons in orbital $\alpha$, one electron
in orbital $\beta$ with spin $\sigma_{\beta}$ and one electron in orbital $\gamma$ with
spin $\sigma_{\gamma}$, fivefold occupation
with four electrons in orbitals $\beta$ and $\gamma$, and one electron
in orbital $\alpha$ with spin $\sigma$, and sixfold occupation, respectively.

The completeness of these boson fields gives rise to the normalization condition as follows:
\begin{eqnarray}
\label{eq:norm}
e^{\dag}_{i}e_{i}+\sum_{\alpha,\sigma}p^{\dag}_{i\alpha\sigma}p_{i\alpha\sigma}
+\sum_{\alpha}b^{\dag}_{i\alpha}b_{i\alpha}
+\sum_{\alpha,\sigma_{\beta},\sigma_{\gamma}}d^{\dag}_{i\alpha\sigma_{\beta}\sigma_{\gamma}}
d_{i\alpha\sigma_{\beta}\sigma_{\gamma}}
+\sum_{\sigma_{\alpha},\sigma_{\beta},\sigma_{\gamma}}r^{\dag}_{i\sigma_{\alpha}
\sigma_{\beta}\sigma_{\gamma}}r_{i\sigma_{\alpha}
\sigma_{\beta}\sigma_{\gamma}}\nonumber\\
+\sum_{\alpha,\beta,\sigma}t^{\dag}_{i\alpha\beta\sigma}t_{i\alpha\beta\sigma}
+\sum_{\alpha}q^{\dag}_{i\alpha}q_{i\alpha}
+\sum_{\alpha,\sigma_{\beta},\sigma_{\gamma}}u^{\dag}_{i\alpha\sigma_{\beta}\sigma_{\gamma}}
u_{i\alpha\sigma_{\beta}\sigma_{\gamma}}
+\sum_{\alpha,\sigma}v^{\dag}_{i\alpha\sigma}v_{i\alpha\sigma}+s^{\dag}_{i}s_{i}=1
\end{eqnarray}
Projecting the original Hamiltonian Eqs. (1) and (2) into the new slave-boson
representation, the multi-orbital Hubbard model Hamiltonian
is described as:
\begin{eqnarray}
\label{eq:Hsb}
H&=&\sum_{i,j,\alpha,\beta,\sigma}t^{ij}_{\alpha\beta}Z^{\dag}_{i\alpha\sigma}Z_{j\beta\sigma}
f^{\dag}_{i\alpha\sigma}f_{j\beta\sigma}
+\sum_{i\alpha\sigma}(\epsilon_{\alpha}-\mu)f^{\dag}_{i\alpha\sigma}f_{i\alpha\sigma}
+U\sum_{i,\alpha}b^{\dag}_{i\alpha}b_{i\alpha}
+J_{H}\sum^{\alpha,\beta,\gamma}_{i,\alpha}(b^{\dag}_{i\beta}b_{i\gamma}+b^{\dag}_{i\gamma}b_{i\beta})\nonumber\\
& &+(U^{\prime}-J_{H})\sum_{i,\alpha,\sigma}d^{\dag}_{i\alpha\sigma\bar{\sigma}}d_{i\alpha\sigma\bar{\sigma}}
+U^{\prime}\sum_{i,\alpha,\sigma}d^{\dag}_{i\alpha\sigma\bar{\sigma}}d_{i\alpha\sigma\bar{\sigma}}
+(U+2U^{\prime}-J_{H})\sum^{\alpha,\beta,\gamma}_{i,\alpha,\sigma}(t^{\dag}_{i\beta\gamma\sigma}t_{i\beta\gamma\sigma}
+t^{\dag}_{i\gamma\beta\sigma}t_{i\gamma\beta\sigma})\nonumber\\
& &+J_{H}\sum^{\alpha,\beta,\gamma}_{i,\alpha,\sigma}(t^{\dag}_{i\beta\alpha\sigma}t_{i\gamma\alpha\sigma}
+t_{i\beta\alpha\sigma}t^{\dag}_{i\gamma\alpha\sigma})
+3(U^{\prime}-J_{H})\sum_{i,\sigma}r^{\dag}_{i\sigma_{\alpha}\sigma_{\beta}\sigma_{\gamma}}r_{i\sigma_{\alpha}\sigma_{\beta}\sigma_{\gamma}}\nonumber\\
& &+(3U^{\prime}-J_{H})\sum^{\alpha,\beta,\gamma}_{i,\alpha,\sigma}r^{\dag}_{a\sigma_{\alpha}\sigma_{\beta}\bar{\sigma_{\gamma}}}
r_{a\sigma_{\alpha}\sigma_{\beta}\bar{\sigma_{\gamma}}}
+2(U+2U^{\prime}-J_{H})\sum_{i,\alpha}q^{\dag}_{i\alpha}q_{i\alpha}
+J_{H}\sum^{\alpha,\beta,\gamma}_{i,\alpha}(q^{\dag}_{i\beta}q_{i\gamma}+q_{i\beta}q^{\dag}_{i\gamma})\nonumber\\
& &+(U+5U^{\prime}-3J_{H})\sum^{\alpha,\beta,\gamma}_{i,\alpha,\sigma}
u^{\dag}_{i\alpha\sigma_{\beta}\sigma_{\gamma}}u_{i\alpha\sigma_{\beta}\sigma_{\gamma}}
+(U+5U^{\prime}-2J_{H})\sum^{\alpha,\beta,\gamma}_{i,\alpha,\sigma}u^{\dag}_{i\alpha\sigma_{\beta}\bar{\sigma_{\gamma}}}
u_{i\alpha\sigma_{\beta}\bar{\sigma_{\gamma}}}\nonumber\\
& &+(2U+8U^{\prime}-4J_{H})\sum_{i,\alpha,\sigma}v^{\dag}_{i\alpha\sigma}v_{i\alpha\sigma}
+(3U+12U^{\prime}-6J_{H})\sum_{i}s^{\dag}_{i}s_{i}
\end{eqnarray}
where the renormalization factor
\begin{eqnarray}
\label{eq:renor}
Z_{i\alpha\sigma}=(1-\tilde{Q}_{i\alpha\sigma})^{-\frac{1}{2}}\tilde{Z}_{i\alpha\sigma}
\tilde{Q}^{-\frac{1}{2}}_{i\alpha\sigma}
\end{eqnarray}
\begin{eqnarray}
\label{eq:renor2}
\tilde{Z}_{i\alpha\sigma}&=&e^{\dag}_{i}p_{i\alpha\sigma}
+p^{\dag}_{i\alpha\bar{\sigma}}b_{i\alpha}
+\sum^{\alpha,\beta,\gamma}_{\sigma^{\prime}}p^{\dag}_{i\beta\sigma^{\prime}}d_{i\gamma\sigma\sigma^{\prime}}
+\sum^{\alpha,\beta,\gamma}_{\sigma^{\prime}}p^{\dag}_{i\gamma\sigma^{\prime}}d_{i\beta\sigma^{\prime}\sigma}
+\sum^{\beta\ne\alpha}_{\beta}b^{\dag}_{i\beta}t_{i\beta\alpha\sigma}\nonumber\\
&&+\sum^{\alpha,\beta,\gamma}_{\sigma^{\prime},\sigma^{\prime\prime}}d^{\dag}_{i\alpha\sigma_{\beta}\sigma_{\gamma}}
r_{i\sigma_{\alpha}\sigma_{\beta}\sigma_{\gamma}}
+\sum^{\alpha,\beta,\gamma}_{\sigma^{\prime}}d^{\dag}_{i\gamma\bar\sigma\sigma^{\prime}}t_{i\alpha\beta\sigma^{\prime}}
+\sum^{\alpha,\beta,\gamma}_{\sigma_{\prime}}d^{\dag}_{i\beta\sigma^{\prime}\bar\sigma}t_{i\alpha\gamma\sigma^{\prime}}
+\sum_{\sigma^{\prime},\sigma^{\prime\prime}}r^{\dag}_{i\bar\sigma_{\alpha}\sigma_{\beta}\sigma_{\gamma}}
u_{i\alpha\sigma_{\beta}\sigma_{\gamma}}\nonumber\\
&&+t^{\dag}_{i\gamma\alpha\bar\sigma}q_{i\beta}
+t^{\dag}_{i\beta\alpha\bar\sigma}q_{i\gamma}
+\sum_{\sigma^{\prime}}t^{\dag}_{i\gamma\beta\sigma^{\prime}}u_{i\gamma\sigma_{\alpha}\sigma_{\beta}}
+\sum_{\sigma^{\prime}}t^{\dag}_{i\beta\gamma\sigma^{\prime}}u_{i\beta\sigma_{\gamma}\sigma_{\alpha}}
+q^{\dag}_{i\alpha}v_{i\alpha\sigma}\nonumber\\
&&+\sum_{\sigma^{\prime}}u^{\dag}_{i\gamma\bar\sigma_{\alpha}\bar\sigma_{\beta}}v_{i\beta\bar\sigma^{\prime}}
+\sum_{\sigma^{\prime}}u^{\dag}_{i\beta\sigma_{\gamma}\bar\sigma_{\alpha}}v_{i\gamma\sigma^{\prime}}
+v^{\dag}_{i\alpha\bar\sigma}s_{i}
\end{eqnarray}
 The corresponding Fermion number constraint for $\alpha$ orbital with spin $\sigma$ reads:
\begin{eqnarray}
\label{eq:pnc}
\tilde{Q}_{i\alpha\sigma}=f^{\dag}_{i\alpha\sigma}f_{i\alpha\sigma}
\end{eqnarray}
where
\begin{eqnarray}
\label{eq:pnc2}
\tilde{Q}_{i\alpha\sigma}&=&p^{\dag}_{i\alpha\sigma}p_{i\alpha\sigma}+b^{\dag}_{i\alpha}b_{i\alpha}
+\sum_{\sigma^{\prime}}d^{\dag}_{i\beta\sigma_{\gamma}\sigma_{\alpha}}d_{i\beta\sigma_{\gamma}\sigma_{\alpha}}
+\sum_{\sigma^{\prime}}d^{\dag}_{i\gamma\sigma_{\alpha}\sigma_{\beta}}d_{i\gamma\sigma_{\alpha}\sigma_{\beta}}
+\sum_{\sigma^{\prime},\sigma^{\prime\prime}}r^{\dag}_{i\sigma_{\alpha}\sigma_{\beta}\sigma_{\gamma}}
r_{i\sigma_{\alpha}\sigma_{\beta}\sigma_{\gamma}}\nonumber\\
& &+\sum_{\beta,\sigma}t^{\dag}_{i\alpha\beta\sigma}t_{i\alpha\beta\sigma}
+\sum_{\beta}t^{\dag}_{i\beta\alpha\sigma}t_{i\beta\alpha\sigma}
+\sum^{\beta\ne\alpha}_{\beta}q^{\dag}_{i\beta}q_{i\beta}
+\sum_{\sigma^{\prime}}u^{\dag}_{i\gamma\sigma_{\alpha}\sigma_{\beta}}u_{i\gamma\sigma_{\alpha}\sigma_{\beta}}
+\sum_{\sigma^{\prime}}u^{\dag}_{i\beta\sigma_{\gamma}\sigma_{\alpha}}u_{i\beta\sigma_{\gamma}\sigma_{\alpha}}\nonumber\\
& &+\sum_{\sigma^{\prime},\sigma^{\prime\prime}}u^{\dag}_{i\alpha\sigma_{\beta}\sigma_{\gamma}}
u^{\dag}_{i\alpha\sigma_{\beta}\sigma_{\gamma}}
+\sum^{\beta\ne\alpha}_{\beta\sigma}v^{\dag}_{i\beta\sigma}v_{i\beta\sigma}
+v^{\dag}_{i\alpha\sigma}v_{i\alpha\sigma}
+s^{\dag}_{i}s_{i}
\end{eqnarray}

Averaging the boson operators in Eq.(4)-(8), we can obtain
an effective mean filed Hamiltonian, hence the total ground-state energy.
The original fermions are
guaranteed by the constraints Eq.(\ref{eq:pnc}),
which are implemented by means of
the corresponding generalized Lagrange multiplier method.
In order to determine the stable magnetic ground state,
we minimize the total energies for different magnetic configurations
based on the pattern search method.
To simplify the calculations, various symmetries should be utilized.
For instance, in the AFM situation with two sublattices,
the single occupation probabilities
$p^B_{\alpha\uparrow}$ in sublattice $B$ are identical
to $p^A_{\alpha\downarrow}$ in sublattice $A$, double
occupation probabilities $d^B_{\alpha\downarrow\downarrow}$
=$d^A_{\alpha\uparrow\uparrow}$, {\it etc.}.

\section{Results and discussions}

In this section, we present main numerical results on the
electronic and magnetic properties within the three-orbital
model for iron selenide systems.
The fillings of three quarters and two thirds, as well as
half-filling, are all considered for comparison.

\subsection{Three-quarter filling case}

We firstly consider the electron filling $\it{n}$=4.5 case,
corresponding to pure KFe$_{2}$Se$_{2}$ compounds.
Note that we adopt the hole representation for convenience within
KRSB mean field approach throughout this paper.
Thus the particle number at the three-quarter filling case is 1.5
within the three-orbital model.
Taking into account several types of magnetic configurations with high symmetry,
the PM, ferromagnetic (FM), $N\acute{e}el$ AFM and SAFM cases,
we find that only the PM phase is the most stable at a filling of three
quarters when $\it{U}$ increases from 0 up to 5 eV.
%
%
\begin{figure}[htbp]\centering
\includegraphics[angle=0, width=0.6 \columnwidth]{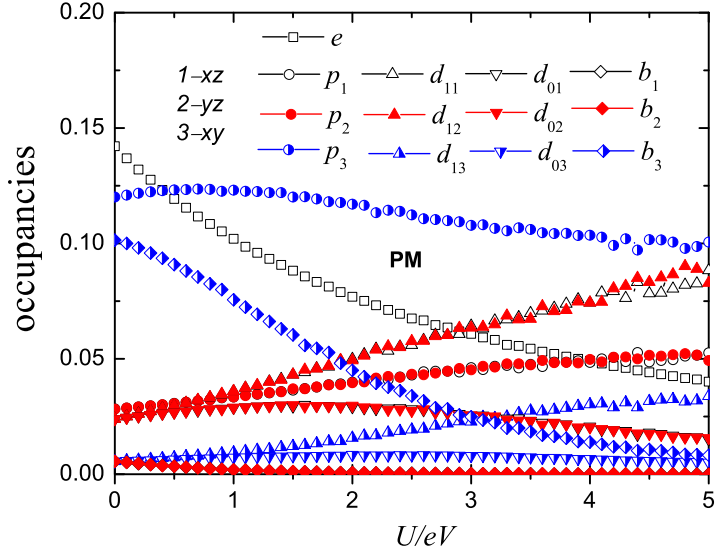}
\caption{Dependence of boson occupancies on Coulomb interaction $\it{U}$ at an
electron filling of $\it{n}$=4.5, and
$\it{J}_{\rm{H}}$=0.25$\it{U}$. Note that in the PM case, $d_{0\alpha}$=
$d_{\alpha\uparrow\downarrow}$=$d_{\alpha\downarrow\uparrow}$, $d_{1\alpha}$=
$d_{\alpha\uparrow\uparrow}$=$d_{\alpha\downarrow\downarrow}$.} \label{fig2}
\end{figure}
%
%
Because of the mixing valence of Fe$^{2+}$ and Fe$^{+}$ in KFe$_{2}$Se$_{2}$,
the system is not a magnetic ordered and insulating state in the homogenous phase.
Since the electron fillings of both Fe$^{2+}$ and Fe$^{+}$ are away from half filling, unlike mixing valent Na$_{0.5}$CoO$_{2}$ \cite{PRL92-247001}, KFe$_{2}$Se$_{2}$ does not form charge ordering.
Previous work on charge susceptibility shows that the
Coulomb interaction suppresses the charge instability \cite{physicab407-1139,PRB84-184521},
thus it is hard to form charge ordering in KFe$_{2}$Se$_{2}$.

\begin{figure}[htbp]
\hspace*{-9mm}
\begin{minipage}[t]{0.45\linewidth}
    \centering
    \includegraphics[width=3.0in]{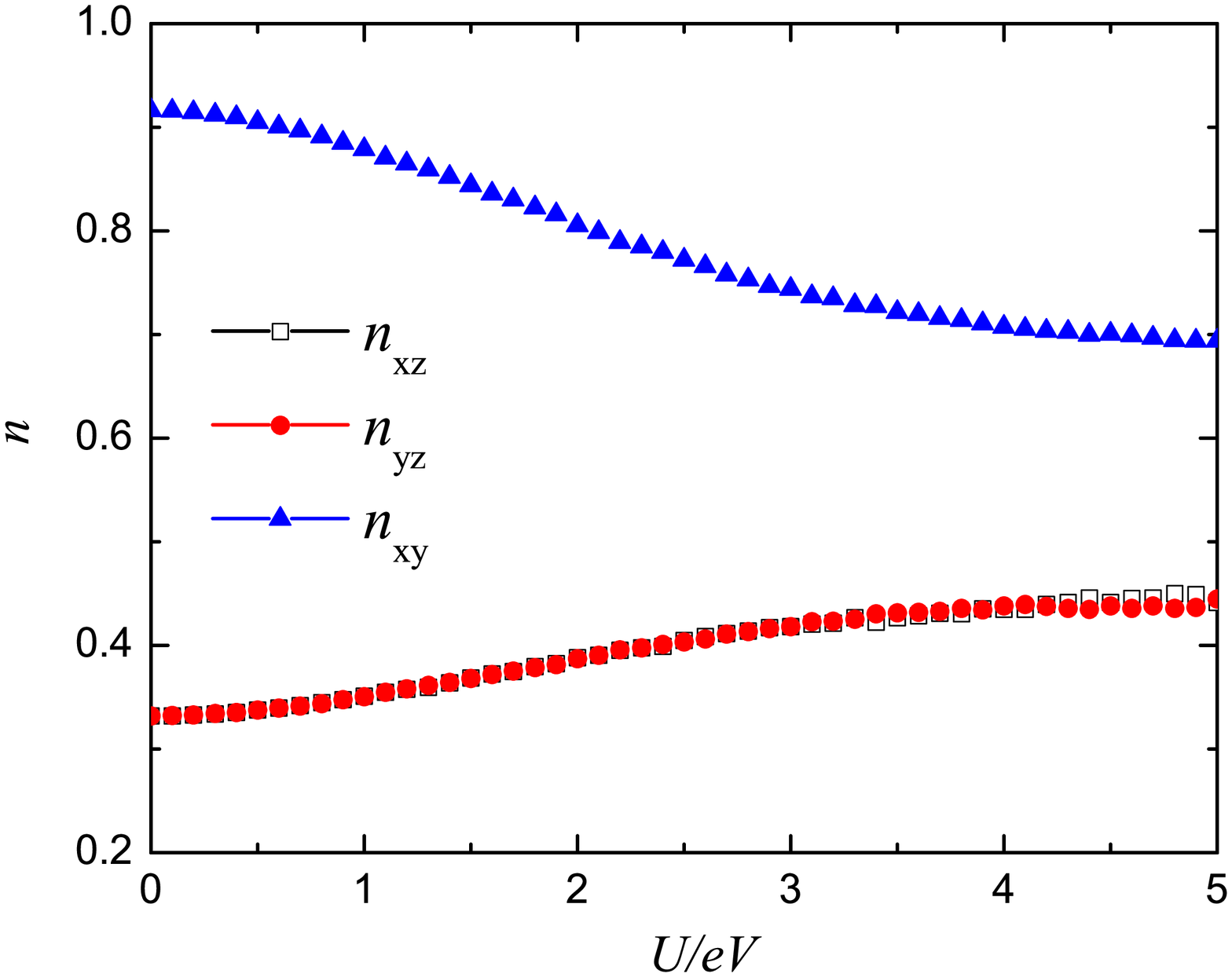}
    \label{Fig3-1}
\end{minipage}
\hspace{0.25ex}
\begin{minipage}[t]{0.45\linewidth}
    \centering
    \includegraphics[width=3.0in]{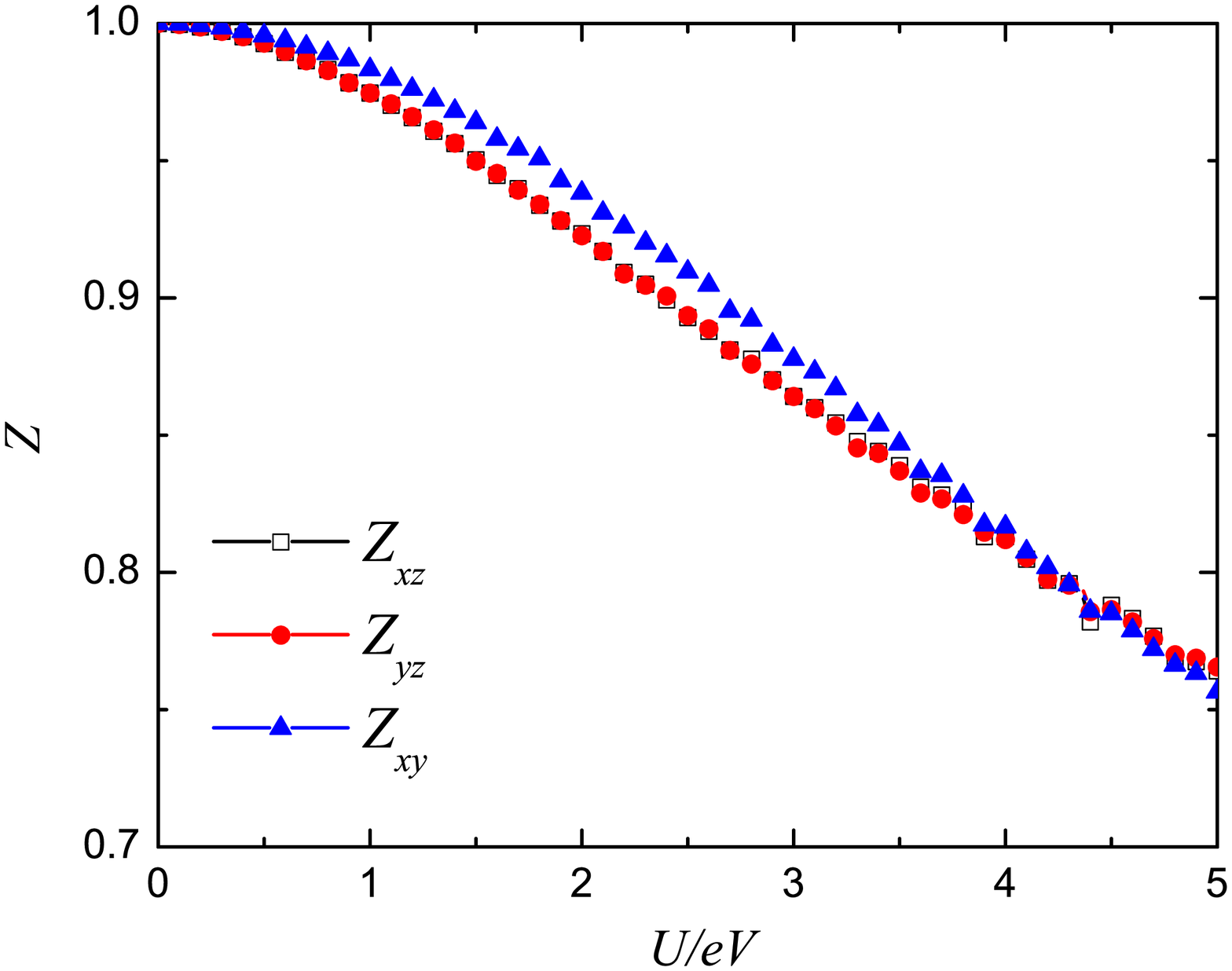}
    \label{Fig3-2}
\end{minipage}
\  \\
\caption{(Color online) Orbital occupations (left panel), and
renormalization factor (right panel) as a function of
the Coulomb interaction $\it{U}$ at $\it{n}$=4.5, and
$\it{J}_{\rm{H}}$=0.25$\it{U}$.}
\end{figure}
Within the present KRSB framework, the dependence of various boson occupancies on
the Coulomb interaction $\it{U}$ at $\it{J}_{\rm{H}}$=0.25$\it{U}$ is plotted in Fig. 2.
It is clearly found that the empty occupation $\it{e}$, single occupation $\it{p}_{3}$ and double occupation $\it{b}_{3}$ with orbital $\it{xy}$ are dominant for small $U$.
With the increase of the Coulomb interaction, the empty occupation $\it{e}$ and
triple occupation $\it{b}_{3}$ decrease sharply, while $\it{d}_{11}$, $\it{d}_{12}$ and $\it{d}_{13}$
($\it{d}_{1\alpha}$=$\it{d}_{\alpha\uparrow\uparrow}$=$\it{d}_{\alpha\downarrow\downarrow}$) increase due to the increasing of the Hund's rule coupling and the Coulomb interaction.
This shows that the system is a PM and the magnetic moment of each Fe spin increases with the lift of $\it{U}$ and $J_{H}$, rather than a nonmagnetic one.
%

\begin{figure}[htbp]\centering
\includegraphics[angle=0, width=0.7 \columnwidth]{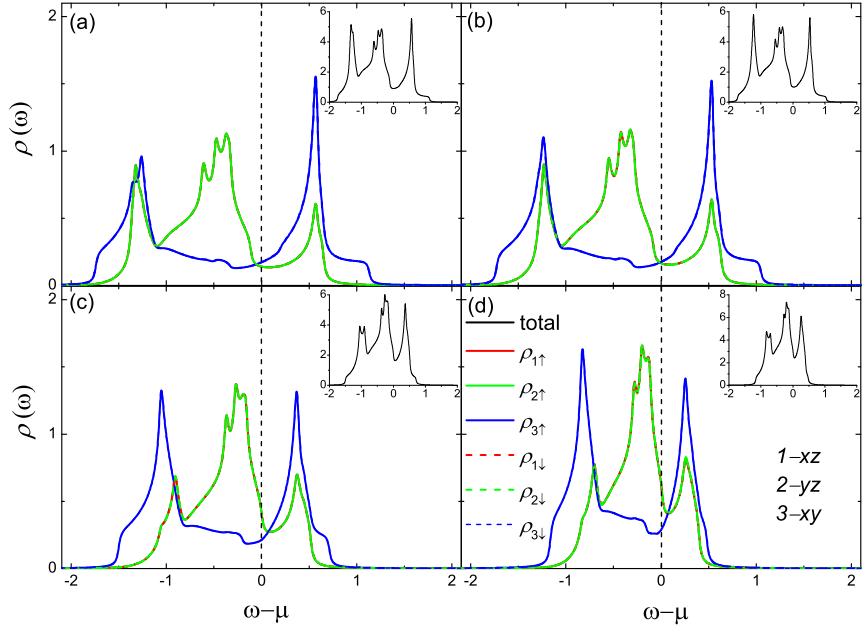}
\caption{Projected densities of states are plotted for $\it{U}$=0 (a),
1.0 eV (b), 3.0 eV (c), and 5.0 eV (d) at $\it{n}$=4.5,
respectively, with $\it{J}_{\rm{H}}$=0.25$\it{U}$. Inset: total densities
of states.
} \label{fig4}
\end{figure}
%
The orbital occupations as a function of the Coulomb interaction
are given in the left panel of Fig. 3.
It is obviously found that the electrons are occupied at the lower
energy $\it{xz}$- and $\it{yz}$-orbitals. And there is no orbital polarization between the
$\it{xz}$- and $\it{yz}$-orbitals. With the increase of Coulomb interaction, a fraction of electrons in the $\it{xz}$- and $\it{yz}$-orbitals transfer to higher energy $\it{xy}$-orbital since larger $J_{H}$ enhances the effect of the Hund's rule. Thus
the high energy $\it{xy}$ orbital occupies more electrons as $J_{H}$ increases.
The renormalization factors of each orbital as a function of the Coulomb interaction $\it{U}$ are shown in the right panel of Fig. 3. With the increase of Coulomb interaction, the bandwidths of the three orbitals become narrower and narrower, and
the renormalization factors become small.
We notice that in the PM phase, the degeneracy of orbital $\it{xz}$ and $\it{yz}$ is not removed, and the renormalization factors of bandwidths in different orbitals are nearly the same. When $\it{U}$ is 3, 4 and 5 eV, the renormalization factor $\it{Z}$ is about 0.85, 0.8 and 0.76, respectively, giving rise to the band mass of original fermions $m_{b}/m=1/Z^{2}$ $\sim$ 1.38, 1.56 and 1.73, respectively, and yielding a mass renormaization factor of about 2 at $\it{U}$=5 eV. We expect that the disorder effect and the spin fluctuations beyond the KRSB mean field approximation will further narrow the bandwidths and enhance the effective mass.
This band narrowing factor is comparable with the experimental ones from ARPES \cite{PRB80-165115} and de Haas-Van Alphen \cite{PRL104-057008},
indicating that KFe$_{2}$Se$_{2}$ lies in the intermediate correlation region,
similar to the FeAs-based systems.
While in the presence of the Fe-vacancy, the measured renormalization factor is about 6.1 in (Tl,Rb)$_{x}$Fe$_{2-y}$Se$_{2}$, greatly larger than 2.
This shows that most probably the insulating properties in K$_{x}$Fe$_{2-y}$Se$_{2}$ systems are induced by the ordering of Fe vacancies, rather than by the strong electronic correlation.

The projected densities of states (PDOS) of KFe$_{2}$Se$_{2}$ for the Coulomb
interaction $\it{U}$=0, 1, 3 and 5 eV are shown in Fig. 4.
The finite DOS at FS shows a pseudo-gap-like structure,
and there is no obvious van Hove singularity.
Nevertheless in FeAs-based compounds, a considerable van Hove
singularities in the DOS is attributed to the FS
nesting with wave vector $\mathbf{Q}$=($\pi$, 0) \cite{PRB77-220503R}.
Therefore, the absence of the van Hove singularities
in the DOS of KFe$_{2}$Se$_{2}$ may be the consequence
of the breakdown of the FS nesting,
which arises from the fact that there
lacks the hole pockets around $\varGamma$ point.
With the increase of the Coulomb interaction from 0 to
5 eV, the bandwidth becomes narrow, $\it{W}$ $\sim$ 2.5 eV at $\it{U}$=3 eV,
smaller than 3.2 eV at $\it{U}$=0, showing
that the system is a correlated bad metal
and is in the intermediate correlation region with $\it{W}$$\sim$$\it{U}$.

Through the analysis above, we find that at 3/4 filling,
the K$_{x}$Fe$_{2-y}$Se$_{2}$ system, corresponding to KFe$_{2}$Se$_{2}$, is a
PM phase. We also clarify that KFe$_{2}$Se$_{2}$ lies in intermediate
correlation region, similar to the FeAs-based compounds.

\subsection{Two-third filling case}

On the other hand, the electron filling at $\it{n}$=4
corresponds to Fe$^{2+}$-based compounds, such as FeSe and LaFeAsO, {\it etc.}.
The dependence of boson occupancy
probabilities on the Coulomb interaction
is also shown in Fig. 5.
Comparing with $\it{n}$=4.5 case, we find that
with the increase of $\it{U}$, the system transits
from a PM metallic phase to a SAFM metallic one at a critical point $\it{U}_{\rm{c}}$$\approx$1.2 eV.
\begin{figure}[htbp]\centering
\includegraphics[angle=0, width=0.6 \columnwidth]{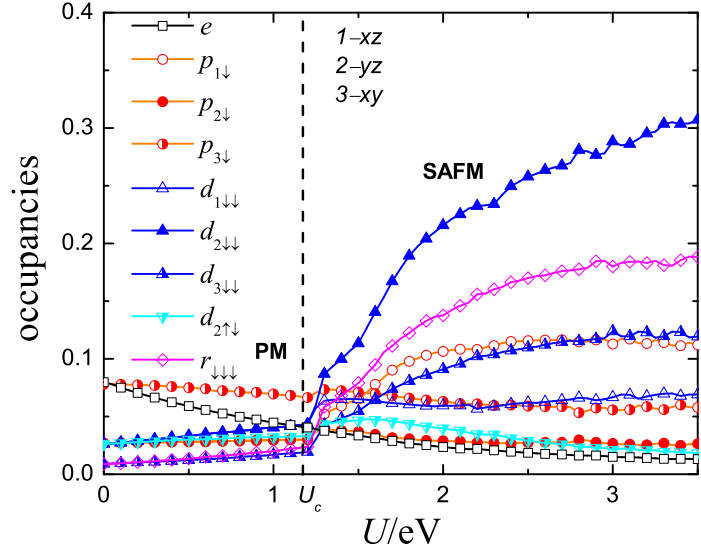}
\caption{Dependence of various boson occupancies on the
Coulomb interaction $\it{U}$ at $\it{n}$=4,
and $\it{J}_{\rm{H}}$=0.25$\it{U}$.} \label{fig5}
\end{figure}
In the PM phase, the single occupations in the $\it{xy}$-orbital $p_{3}$
and empty occupancy {\it e} are dominant. Other multiple occupation states also contribute small but finite weights. While in the SAFM phase, the single,
double and triple occupations with the same spin alignment, $\it{p}$, $\it{d}$ and $\it{r}$, are dominant.
The spin singlet and small spin states contribute very little.
With increasing the Hund's rule coupling, the empty and single
occupations continuously decrease, while the double and triple ones considerably increase.
This behavior arises from the fact that the increase of Coulomb interaction favors the formation of the SAFM phase, and large Hund's rule coupling favors large spin state.
%

The gap opening behavior of the SDW states in the multi-orbital
FeAs-based system is an interesting but unsolved topic.
In order to resolve the behavior of the SDW
gap opening in K$_{x}$Fe$_{2-y}$Se$_{2}$, we present the band dispersions of the PM
and SDW states in Fig. 6.
\begin{figure}[htbp]\centering
\includegraphics[angle=0, width=0.6 \columnwidth]{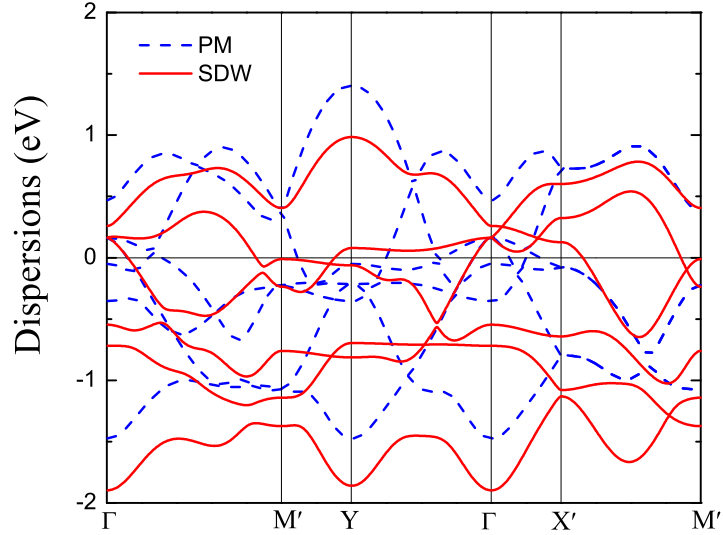}
\caption{Band dispersions of the PM and SDW states along the high symmetry points, $\varGamma$ (0,0), $Y$ (0, $\pi$), $M'$ ($\pi$/2,$\pi$), $X'$ ($\pi$/2,0), in the folded BZ with $\it{U}$=0 and 2 eV, and $\it{J}_{\rm{H}}$=0.25$\it{U}$ at $\it{n}$=4.} \label{fig6}
\end{figure}
%
We find that in the SDW states, only partial SDW gaps open
near $E_{F}$ in comparison with PM states, which is
the consequence of the the band narrowing and spin splitting, similar to FeAs-based
compounds. The partial opening of SDW gap indicates
the system is a bad metal.

\begin{figure}[htbp]
\hspace*{-9mm}
\begin{minipage}[t]{0.42\linewidth}
    \centering
    \includegraphics[width=3.0in]{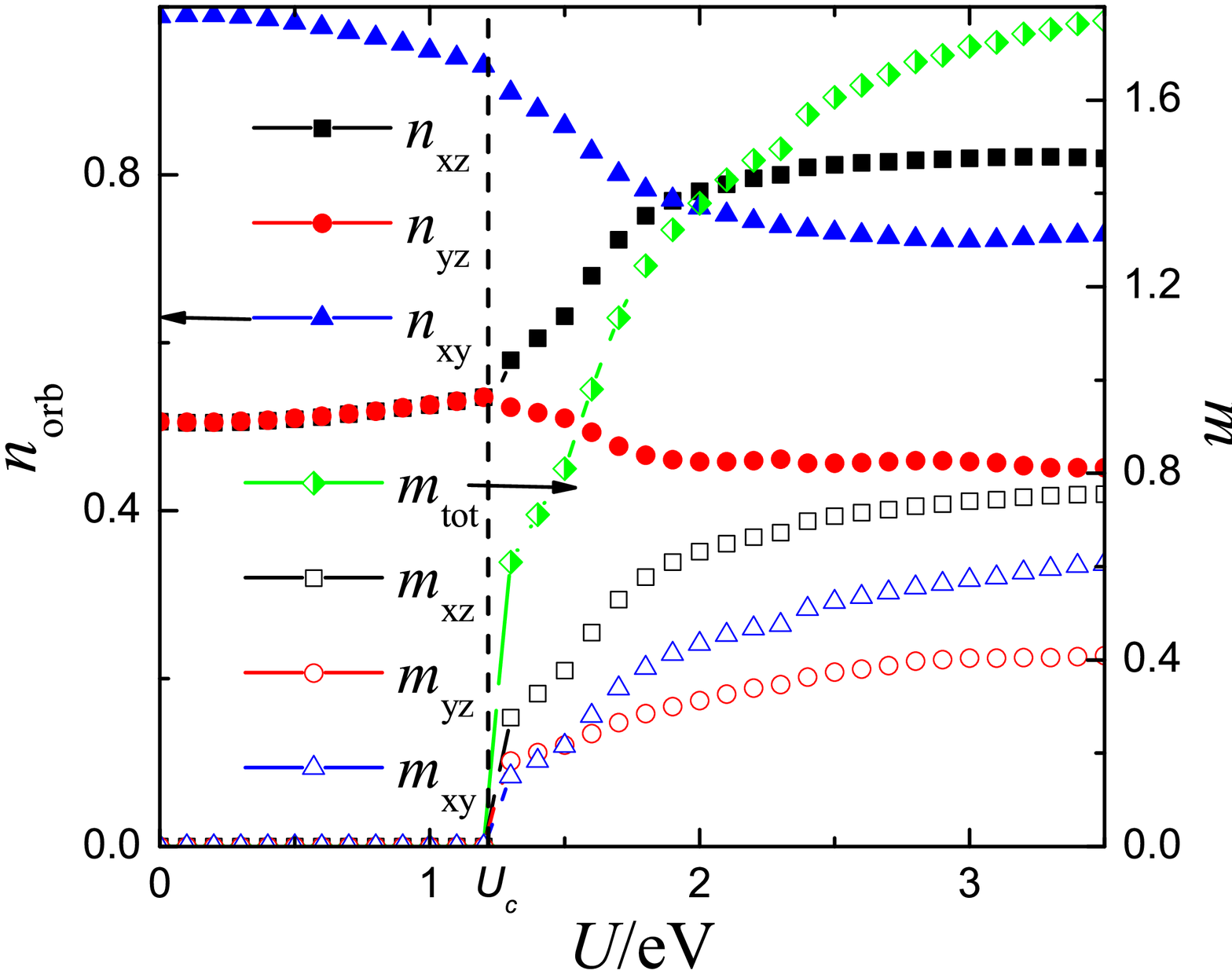}
    \label{Fig7-1}
\end{minipage}
\hspace{0.25ex}
\begin{minipage}[t]{0.42\linewidth}
    \centering
    \includegraphics[width=3.0in]{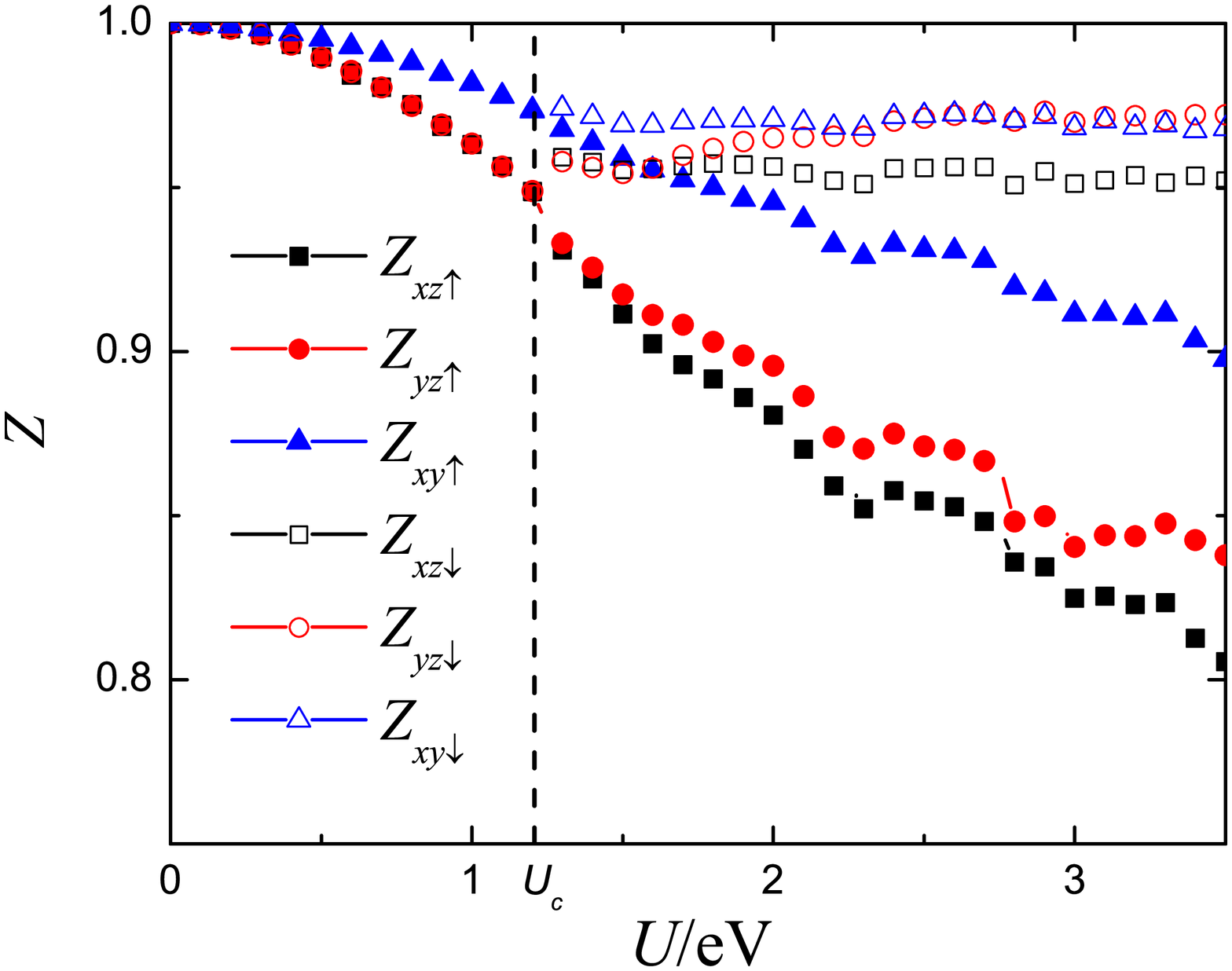}
    \label{Fig7-2}
\end{minipage}
\  \\
\caption{(Color online) Dependence of orbital occupations and magnetic moment
of each orbital (left panel), and
renormalization factor (right panel) on the
Coulomb interaction $\it{U}$ at $\it{n}$=4, and $\it{J}_{\rm{H}}$=0.25$\it{U}$.}
\end{figure}
The dependence of orbital occupations and magnetic moment of each orbital on the Coulomb
interaction $\it{U}$ is also shown in Fig. 7. With the increase of $\it{U}$,
once the system enters the SAFM phase, there is obvious orbital
polarization between $\it{xz}$ and $\it{yz}$ orbitals together with $\it{n_{xz}}< \it{n_{yz}}$ in the electronic representation. And the magnetic moments on the $\it{xz}$ and $\it{yz}$ orbitals are different with $m_{xz}>m_{yz}$.
These are consistent with the previous LaFeAsO results \cite{PRB84-064435,JPCM24-085603}.
The renormalization factors $\it{Z}$ displayed in the right panel of Fig. 7 show that
the $\it{Z}_{\it{xz}\uparrow}$, $\it{Z}_{\it{yz}\uparrow}$ and $\it{Z}_{\it{xy}\uparrow}$ become smaller and smaller with $\it{Z}_{\it{xz/yz}\uparrow}$$<$ $\it{Z}_{\it{xy}\uparrow}$ when $\it{U}$ increases. This indicates that the SDW gap opening mainly occurs in the $\it{xz}$/$\it{yz}$ orbitals.

\begin{figure}[htbp]\centering
\includegraphics[angle=0, width=0.7 \columnwidth]{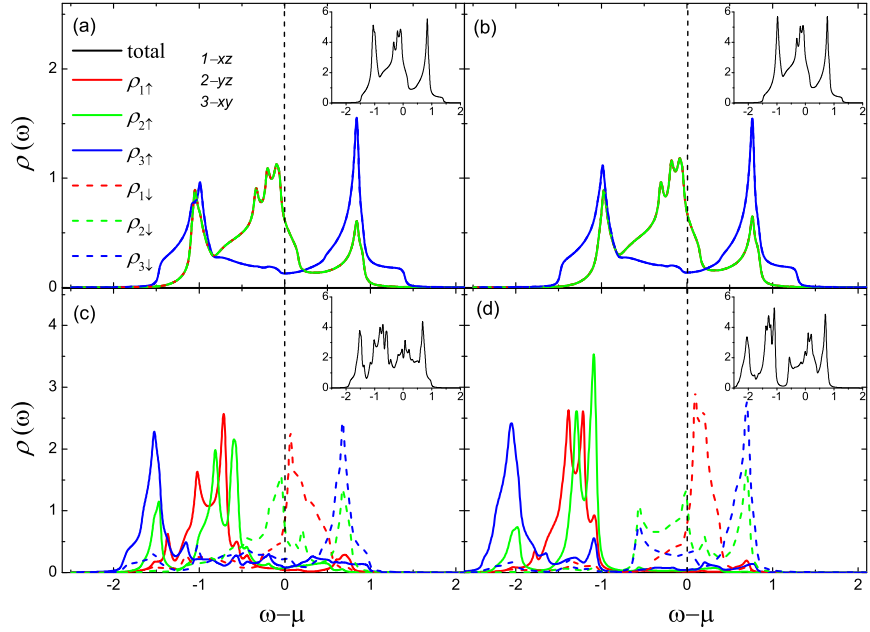}
\caption{Projected densities of states are plotted for $\it{U}$=0 (a),
1.0 eV (b), 2.0 eV (c), and 3.0 eV (d) at $\it{n}$=4, respectively,
with $\it{J}_{\rm{H}}$=0.25$\it{U}$. Inset: total densities
of states.} \label{fig8}
\end{figure}
The evolution of the PDOS with Coulomb interactions $\it{U}$ is also displayed in Fig. 8.
We find that at $\it{U}$=2.0 eV and 3.0 eV, there exhibit obvious spin polarizations in SAFM phase due to the breakdown of spin symmetry. The $\it{xz}$- and $\it{yz}$- orbitals dominate the FS, while the $\it{xy}$ orbital mainly lies far from the Fermi levels. This also shows that the $\it{xz}$/$\it{yz}$ orbitals determine the electronic properties near $E_{F}$, and mainly involve with the formation of the SDW states, consistent with the result of renormalization factor in Fig. 7.

Since the electronic correlation is intermediate, one expects that $U$ is larger than 2 eV in
iron pnictides and selenides; thus at 2/3 filling, the SAFM state with orbital ordering is stable over a wide electron correlation range, from intermediate to strong correlations, which is similar to the FeAs-based systems.

\subsection{Half-filling case}

When turning to the half-filling case, which corresponds to the Fe$^{3+}$-based systems,
such as KFeSe$_{2}$ and FeAs compounds, {\it etc.}, we find that the phase diagram becomes
much richer.
It is found that there exist three critical points when $U$ increases: the system
transits from a PM metal to an SAFM metal at $U_{c_{1}}$, from an SAFM metal to a
$N\acute{e}el$ AFM metallic phase at $U_{c_{2}}$, and from an
AFM metal with intermediate-spin state to an AFM OSMP with high-spin state at $U_{c_{3}}$,
as shown in Fig. 9. These phases will be addressed in detail in the following.
In comparison with $\it{n}$=4.5 and 4 cases, we find that besides the PM and SAFM phases, the $N\acute{e}el$ AFM metallic phase appears in slightly large $\it{U}$ region at $U_{c_{2}}$$\approx$0.8 eV. Only when in the narrow Coulomb interaction region at $U_{c_{1}}$$\approx$0.65$<$$U$$<$ $U_{c_{2}}$, the SAFM phase is stable, as the dependence of the boson occupancies on Coulomb interaction $\it{U}$ showing in Fig. 9.
\begin{figure}[htbp]\centering
\includegraphics[angle=0, width=0.6 \columnwidth]{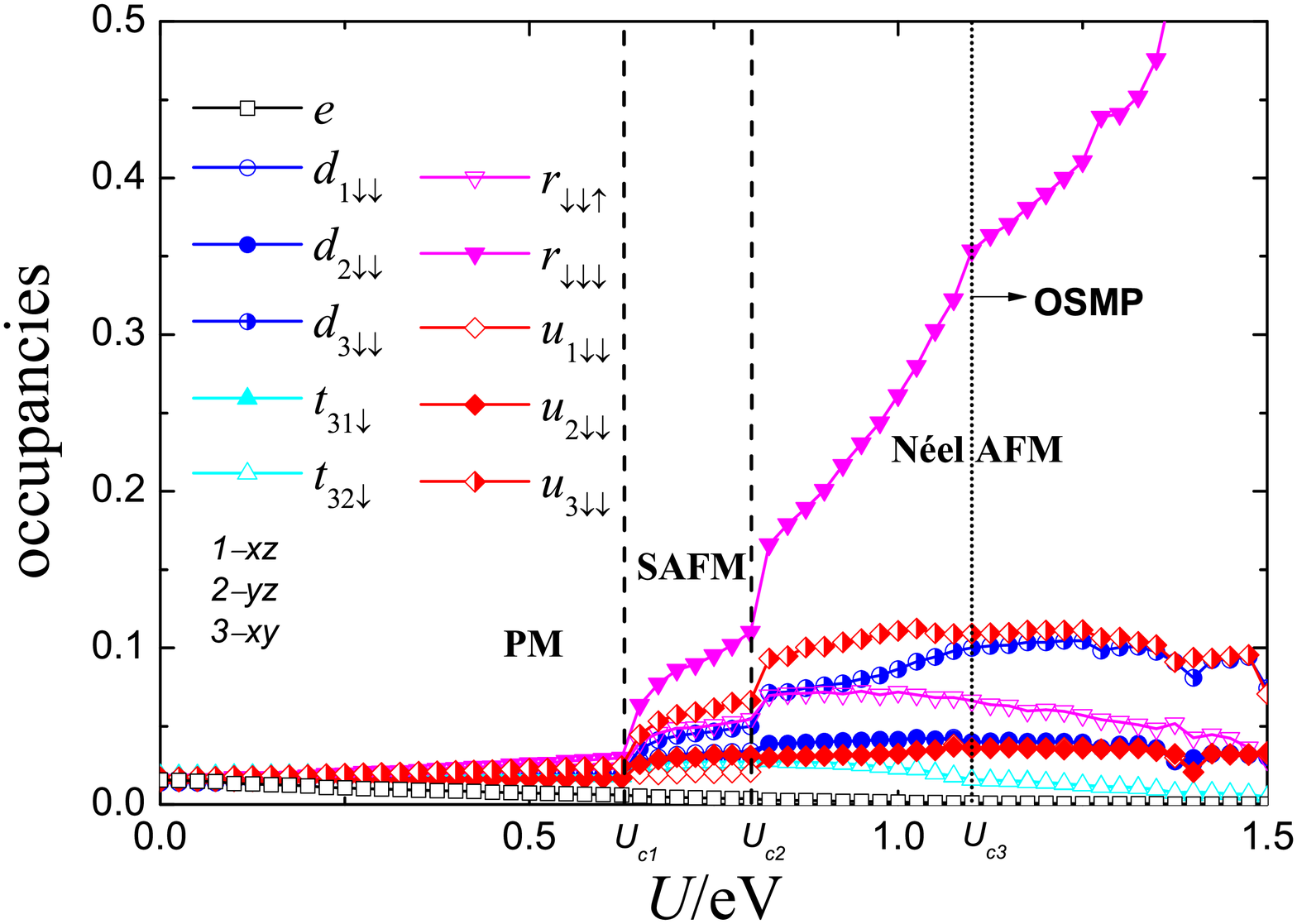}
\caption{Dependence of boson occupancies on Coulomb interaction
$\it{U}$ at half-filling $\it{n}$=3 and
$\it{J}_{\rm{H}}$=0.25$\it{U}$. Three dashed vertical lines denote the
phase boundaries of the PM-SAFM, SAFM-$N\acute{e}el$ AFM, and
$N\acute{e}el$ AFM-OSMP transitions at $U_{c_{1}}$, $U_{c_{2}}$ and $U_{c_{3}}$,
respectively.} \label{fig9}
\end{figure}
%
With the increase of the Coulomb interaction, the triple occupation $\it{r}_{\downarrow\downarrow\downarrow}$ sharply increases, while
the other high occupations are relatively small. This shows that the system
undergoes a spin state transition with the increase of the Coulomb interaction
and the Hund's rule coupling, as we see the low-spin ($S$=1/2) to
intermediate-spin ($S$=1) transition at $U_{c_{2}}$ in Fig. 10.

\begin{figure}[htbp]
\hspace*{-9mm}
\begin{minipage}[t]{0.5\linewidth}
    \centering
    \includegraphics[width=3.45in]{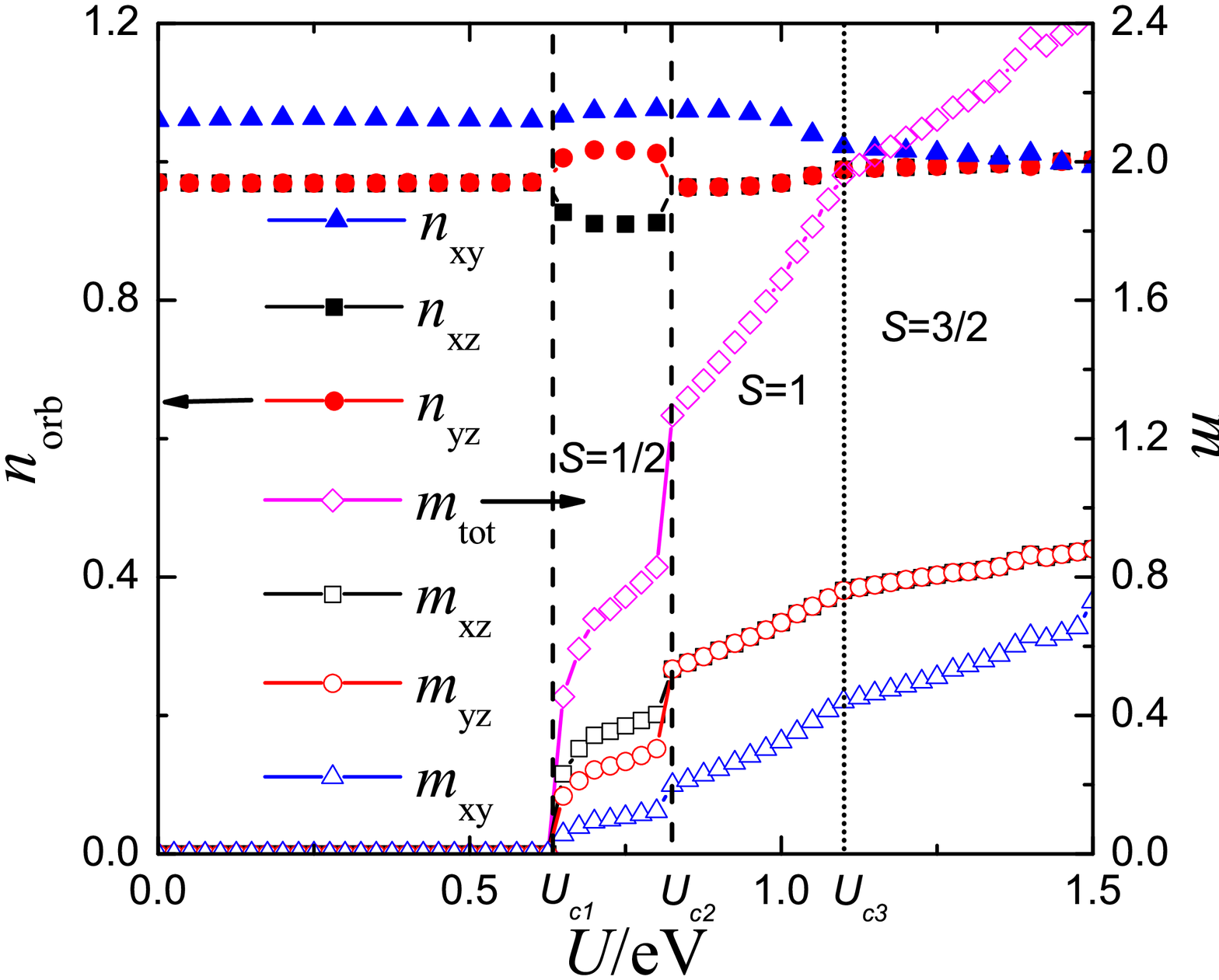}
    \label{Fig10-1}
\end{minipage}
\hspace{0.25ex}
\begin{minipage}[t]{0.5\linewidth}
    \centering
    \includegraphics[width=3.45in]{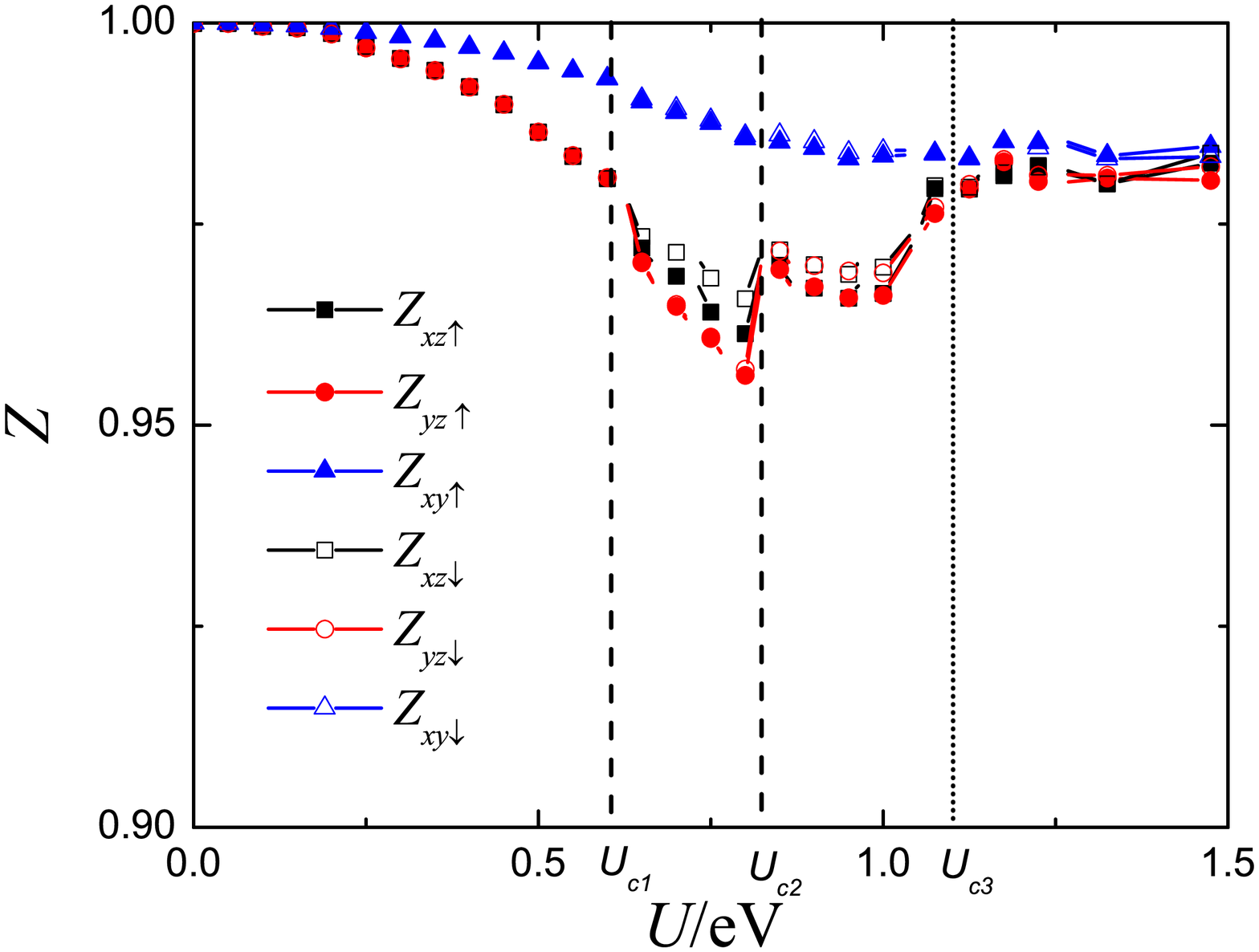}
    \label{Fig10-2}
\end{minipage}
\  \\
\caption{(Color online) Orbital occupations and magnetic moment of each orbital
(left panel), and renormalization factor (right panel) as a function of Coulomb
interaction $\it{U}$ at half-filling $\it{n}$=3, and $\it{J}_{\rm{H}}$=0.25$\it{U}$.
The dashed lines indicate the phase boundaries $U_{c_{1}}$, $U_{c_{2}}$ and $U_{c_{3}}$}
\end{figure}

On the other hand, the dependence of the orbital occupations and the magnetic
moments of three orbitals on Coulomb interaction $\it{U}$ is plotted in the left
panel of Fig. 10. We find that in the SAFM phase, a small orbital polarization appears
with $\it{n_{yz}}>\it{n_{xz}}$ in the present hole representation,
{\it i.e.} $\it{n_{xz}}>\it{n_{yz}}$ in the electron representation.
This supports the itinerant orbital ordering in the parent phases of iron pnictides \cite{PRB84-064435,EPJB85-55}.
Meanwhile the magnetic moments on different orbitals possess $\it{m_{yz}}<\it{m_{xz}}$
with total magnetic moment $m_{tot}$$<$1$\mu_{B}$. The system lies in a low-spin state.
However, in the $N\acute{e}el$ AFM metallic phase, there is no orbital polarization, since in the presence of the spin-orbital coupling, the preserved spin rotational symmetry does not lift the orbital degeneracy or break the orbital symmetry.
When $\it{U}$$>$$U_{c_{3}}$, the system enters the OSMP phase, as displayed in the left panel of Fig. 10. All of the three orbital occupations are nearly equal to 1. The magnetic moment per orbital steeply increases, and the total magnetic moment $m_{tot}$ is larger than 2$\mu_{B}$. Meanwhile, the system undergoes an intermediate-spin ($S$=1) to high spin ($S$=3/2) transition with the increasing of $U$, and thus enters a high-spin state when $\it{U}$$>$$U_{c_{3}}$.

The right panel of Fig. 10 displays the renormalization factor of each orbital
as a function of Coulomb interaction $\it{U}$, which is in sharp contrast with
the $\it{n}$=4.5 and 4 cases.
In the PM and SAFM phases, all the renormalization factors $Z_{xz}$, $Z_{yz}$ and $Z_{xy}$
($Z_{xz/yz}$$<$$Z_{xy}$) smoothly decrease with increasing $U$, indicating the
bandwidths become narrow due to the increase of the Coulomb correlation.
When $\it{U}$$>$$U_{c_{2}}$, the system enters the $N\acute{e}el$ AFM state.
In this situation $Z_{xy}$ gradually decreases, while $Z_{xz}$ and $Z_{yz}$ considerably
change with $U$, suggesting that the variations of orbital and magnetic states mainly occur
in these two orbitals.
The lift of $Z_{xz}$ and $Z_{yz}$ implies the bandwidths of the two orbitals
anomalously increase, which is attributed to the fact that the exchange splitting
of spin-up and spin-down subbands of the $xz$- and $yz$-orbitals increases with
the increase of $U$, leading to the total bandwidth broadening, which is also
seen in Fig. 11 in what follows.

\begin{figure}[htbp]\centering
\includegraphics[angle=0, width=0.7 \columnwidth]{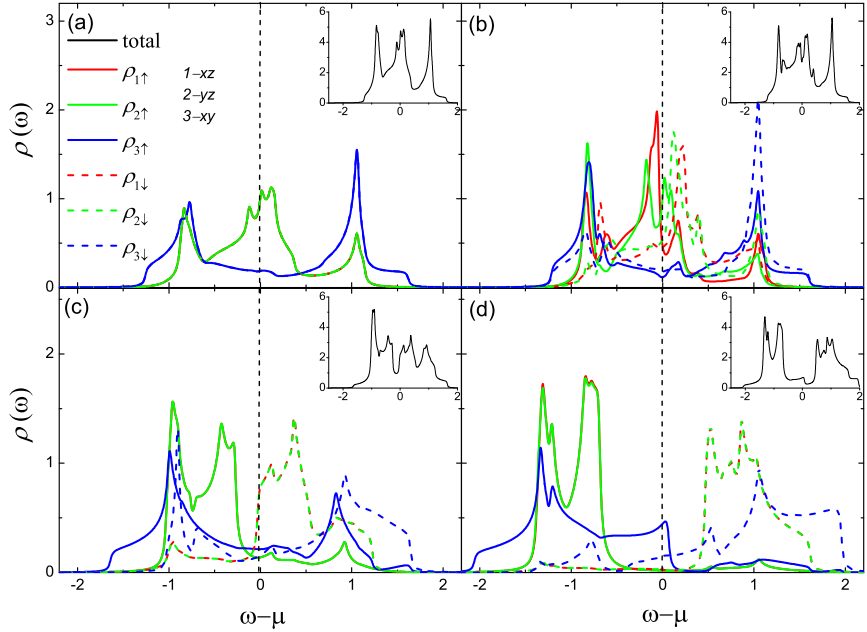}
\caption{Projected densities of states are plotted for $\it{U}$=0 (a),
0.75 eV (b), 1.0 eV (c), and 1.5 eV (d) at
half-filling $\it{n}$=3, respectively.
$\it{J}_{\rm{H}}$=0.25$\it{U}$. Inset: total densities
of states.} \label{fig11}
\end{figure}

We present the PDOS of four typical phases, including PM, SAFM, $N\acute{e}el$
AFM metal and AFM OSMP, in Fig. 11 for Coulomb interaction
$\it{U}$=0, 0.75, 1.0 and 1.5 eV, respectively. It is found that in relatively
small $\it{U}$ region, {\it i.e.} $\it{U}$$\le$ $U_{c_{2}}$,
the $\it{xz}$/$\it{yz}$ orbitals dominate the FS in both the PM and SAFM phases.
Different from Fig. 11(a), in the SAFM phase with $\it{U}$=0.75 eV in Fig. 11(b),
the orbital degeneracy between $\it{xz}$ and $\it{yz}$ orbitals is lifted, consistent
with the orbital polarization in the left panel of Fig. 10.
Meanwhile in the $N\acute{e}el$ AFM metallic state with $\it{U}$=1.0 eV in Fig. 11(c),
the weight of the $\it{xz}$ and $\it{yz}$
orbitals in the FS is greatly suppressed; but the spin splitting becomes large,
consistent with the intermediate spin configuration in Fig. 10.
Interestingly, we find that when $U$$>$$\it{U_{c_{3}}}$$\approx$1.1 eV, an orbital-selective
Mott transition occurs, {\it i.e.} an OSMP emerges. In this situation, only the
broad $\it{xy}$ orbital contributes to the FS, while the narrow $\it{xz}$ and $\it{yz}$
orbitals are insulating and sank below the FS, as seen the PDOS with $U$=1 eV
in Fig. 11(d).

Therefore, at a half filling, the system is mainly a
$N\acute{e}el$ AFM state without orbital ordering in
the intermediate and strong correlations, in addition to an OSMP phase
related with an intermediate-spin to high-spin transition. Compared
with the other band fillings, the different magnetic phase
diagrams suggest a band-filling controlling magnetism scenario in the
iron selenide systems.

\section{Discussion and Summary}

The scenario of the orbital selective Mott transition in K$_{x}$Fe$_{2-y}$Se$_{2}$
could be understood in a sketch of OSMP in the presence of magnetism shown in Fig. 12
through tuning the band filling and the correlation strength.
When tuning to the half-filling, with the increase of Coulomb interaction and Hund's rule coupling, the spin-up and spin-down subbands split with each other, and the spin-up $\it{xz}$- and $\it{yz}$-orbitals  are filled, sank below $E_F$ and become insulating; while the $\it{xy}$-orbital remains across $E_F$ and is conducting, giving rise to an AFM OSMP.
We notice that in comparison with the bandwidths of K$_{x}$Fe$_{2-y}$Se$_{2}$, the $U_{c_{3}}$ for the occurrence of the OSMP is small. This may arise from the following two
reasons: (1) in the parent phase of Fe-based superconductors, the Hund's rule coupling is large with $J_{H}$=0.25$\it{U}$.
If small values $J_{H}$=0.15$\it{U}$ or 0.1$\it{U}$ are adopted, $U_{c_{3}}$ will
reach a relatively large values up to 2 or even 3 eV; (2) $U_{c_{3}}$ is obtained
for the OSMP with AFM order in the present study, it will become larger in the PM situation.
Since KFeSe$_{2}$ possesses not only tetrahedra crystal field but also large
magnetic moment, it may be a potential candidate for such an OSMP phase.
We expect that further increasing the Coulomb interaction $\it{U}$ will
drive the system to a magnetic insulating state accompanied
with a metal-insulator transition.
\begin{figure}[htbp]\centering
\includegraphics[angle=0, width=0.5 \columnwidth]{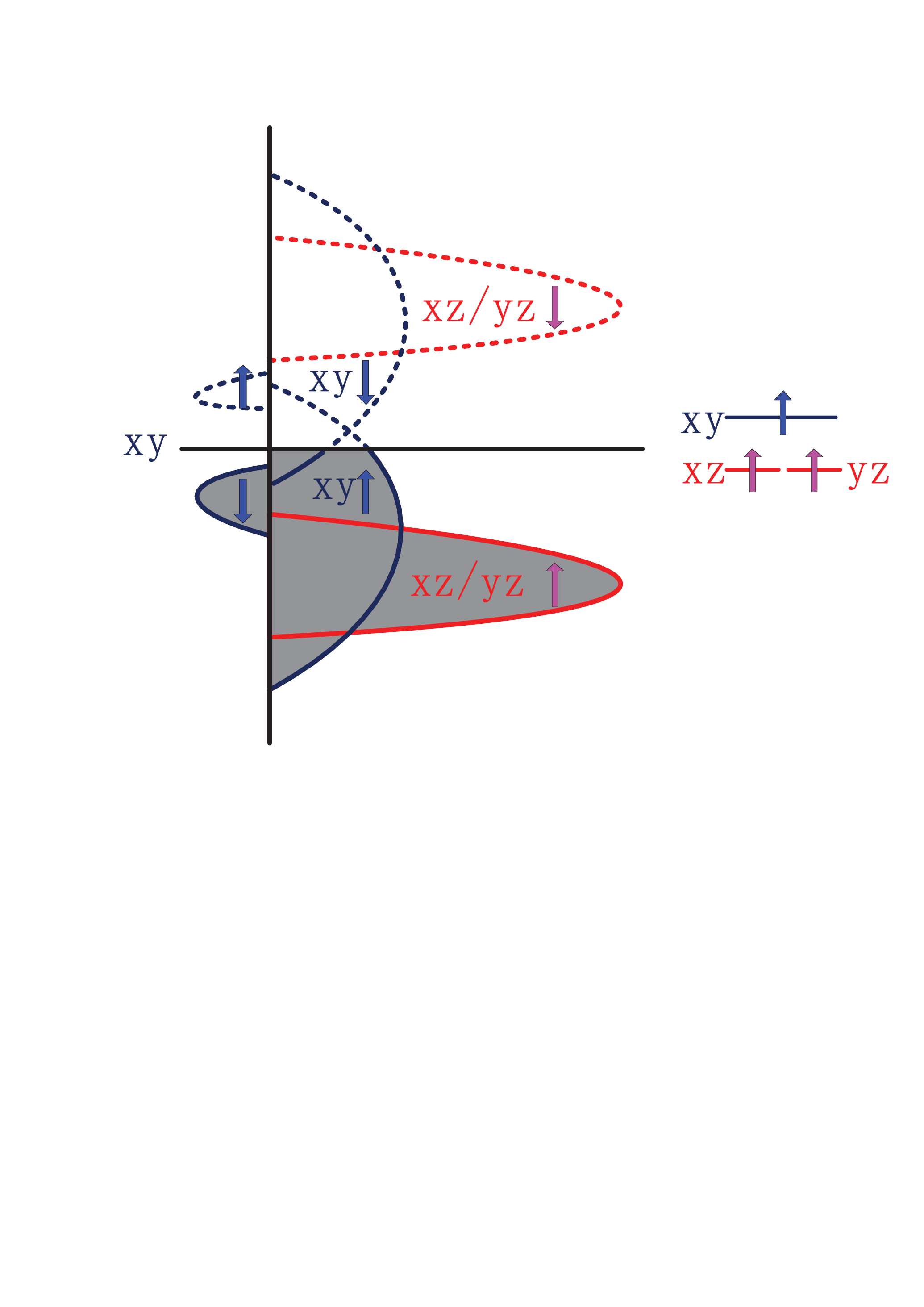}
\caption{Sketch of sublattice band structures in the orbital-selective Mott phase with $N\acute{e}el$ antiferromagnetism for half-filled three-orbital model in K$_{x}$Fe$_{2-y}$Se$_{2}$.
} \label{fig12}
\end{figure}

%
We notice that different from our PM results, the first-principles electronic structure
calculations suggested the ground state in KFe$_{2}$Se$_{2}$ is the SAFM ordering
\cite{CPL28-057402}, or the bi-collinear AFM ordering
resulting from the interplay among the nearest, the next-nearest, and the
next-next-nearest-neighbor superexchange interactions mediated by Se
4$\it{p}$-orbitals in AFe$_{2}$Se$_{2}$ (A=K, Tl, Rb, or Cs) \cite{PRB84-054502}.
In addition, some other iron-based materials, such as
FeSe \cite{PRL102-177003,PRL102-177005}, LiFeAs \cite{nmat10-932},
and KFe$_{2}$As$_{2}$ \cite{PRL103-047002,PRL106-067003,nmat10-932}, {\it etc.},
have no magnetism at all as observed in the experiments, but AFM is obtained in the
LDA calculations.
The discrepancies among these materials suggest the magnetism is
sensitive to the electronic properties or lattice distortion \cite{nmat10-932}.
The LDA methods usually so overestimate the magnetic moment and AFM ordering
but omit some spin fluctuations of the system due to the intermediate
electronic correlation that a proper treatment on the electronic correlation
in these FeSe-based compounds may be important for understanding its magnetic
ground state.
In our study, we deal with the electronic correlation within
the framework of the KRSB approach which is verified as an effective
approach to treat the electronic correlation ranging from weak
through intermediate to strong ones.
Our PM groundstate result, not non-magnetic one, indicates that there exists
strong spin fluctuation which is regarded as the superconducting
pairing mechanism, thus KFe$_{2}$Se$_{2}$ is a potential candidate of
the parent phase for superconductor without doping.
Moreover, a recent STS experiment demonstrated the existence
of phase separation and distinguished the contributions: the
KFe$_{2}$Se$_{2}$ component contributes to superconductivity,
while the K$_{0.8}$Fe$_{1.6}$Se$_{2}$ Fe-vacancy ordering
component to the insulating properties \cite{nphys8-126}, suggesting
that pure KFe$_{2}$Se$_{2}$ is more possibly a PM phase when in the
normal state. These observations are consistent with our present results.

In summary, starting with an effective three-orbital model for
newly found KFe$_{2}$Se$_{2}$, we have shown that the ground state
of KFe$_{2}$Se$_{2}$ at three-quarter filling is a paramagnetic metallic phase.
We also suggest a possible OSMP phase through tuning the band-filling
in K$_{x}$Fe$_{2-y}$Se$_{2}$.
Our results demonstrate that the band filling plays a
key role in the electronic and magnetic properties of K$_{x}$Fe$_{2-y}$Se$_{2}$.
Since the phase separation widely exists in K$_{x}$Fe$_{2-y}$Se$_{2}$
materials, future experiments are expected to confirm
the magnetic ground state of KFe$_{2}$Se$_{2}$.
In addition, the influences of the band filling and correlation on the electronic
structures and magnetic properties in the presence of Fe vacancies for
K$_{x}$Fe$_{2-y}$Se$_{2}$ is an interesting topic, thus further theoretical
works are expected.

\acknowledgements
The author (D.Y.) gratefully acknowledge the help of Da-Mei Zhu with the manuscript.
This work was supported by the Natural Science
Foundation of China (NSFC) No. 11104274, 11074257, and of Anhui
Province No. 11040606Q56, and the Knowledge Innovation Program of
the Chinese Academy of Sciences. Numerical calculations were
performed at the Center for Computational Science of CASHIPS.

\bibliography{apsrev4-1}

\end{document}